\documentclass[journal]{IEEEtranTCOM}

\usepackage[T1]{fontenc}
\usepackage[english]{babel}	
\usepackage[latin1]{inputenc}

\usepackage{amsmath,
	amssymb,
	amsthm}

\usepackage{balance}
\usepackage[noadjust]{cite}

\usepackage{color}
\usepackage{graphicx}
\graphicspath{{figures/}}

\usepackage{pgfplots}
\pgfplotsset{compat=1.12,
	legend style={font=\footnotesize},
}

\colorlet{green}{green!60!black}
\IEEEoverridecommandlockouts

\newtheorem{example}{Example}
\newtheorem{lemma}{Lemma}

\newtheorem{theorem}{Theorem}

\usetikzlibrary{calc}
\tikzset{
	partial ellipse/.style args={#1:#2:#3}{
		insert path={+ (#1:#3) arc (#1:#2:#3)}
	}
}

\renewcommand{\d}{~\mathrm{d}}
\DeclareMathOperator*{\E}{\mathbb{E}}

\newcommand{\N}{\mathbb{N}}
\newcommand{\R}{\mathbb{R}}

\newcommand\ceil[1]{\lceil#1\rceil}
\newcommand\floor[1]{\lfloor#1\rfloor}
\newcommand{\cdf}[2]{F_{#1}(#2)}
\newcommand{\e}{\mathrm{e}}

\newcommand{\file}{i}

\newcommand{\numsbs}{B}

\newcommand{\ratecache}{R_\text{C}}
\newcommand{\ratembs}{R_\text{MBS}}
\newcommand{\ratesbs}{R_\text{SBS}}
\newcommand{\ratetilde}{\tilde{R}_\text{SBS}}

\newcommand{\cost}{\theta}
\newcommand{\updatecost}{\cost_\text{C}}
\newcommand{\mbscost}{\cost_\text{MBS}}
\newcommand{\sbscost}{\cost_\text{SBS}}

\newcommand{\rmbs}{r_\text{MBS}}
\newcommand{\rsbs}{r_\text{SBS}}

\title{Dynamic Coded Caching in Wireless Networks}

\begin{document}

\author{Jesper Pedersen, Alexandre Graell i Amat,~\IEEEmembership{Senior Member,~IEEE}, Jasper Goseling,~\IEEEmembership{Member,~IEEE},\\ Fredrik Br\"annstr\"om,~\IEEEmembership{Member,~IEEE}, Iryna Andriyanova,~\IEEEmembership{Member,~IEEE}, and Eirik Rosnes,~\IEEEmembership{Senior Member,~IEEE}
\thanks{This work was funded by the Swedish Research Council under grant 2016-04253 and by the National Center for Scientific Research in France under grant CNRS-PICS-2016-DISCO.}
\thanks{J. Pedersen, A. Graell i Amat, and F. Br\"annstr\"om are with the Department of Electrical Engineering, Chalmers University of Technology, SE-41296 Gothenburg, Sweden (e-mail: \{jesper.pedersen, alexandre.graell, fredrik.brannstrom\}@chalmers.se).}
\thanks{J. Goseling is with the Department of Applied Mathematics, University of Twente, 7522 Enschede, The Netherlands (e-mail: j.goseling@utwente.nl).}
\thanks{I. Andriyanova is with the ETIS-UMR8051 group, CY Cergy Paris University/ENSEA/CNRS, France (e-mail: iryna.andriyanova@ensea.fr).}
\thanks{E. Rosnes is with Simula UiB, N-5020 Bergen, Norway (e-mail: eirikrosnes@simula.no).}
\vspace{-3ex}}

\maketitle

\begin{abstract}
We consider distributed and dynamic caching of coded content at small base stations (SBSs) in an area served by a macro base station (MBS). Specifically, content is encoded using a maximum distance separable code and cached according to a time-to-live (TTL) cache eviction policy, which allows coded packets to be removed from the caches at periodic times. Mobile users requesting a particular content download coded packets from SBSs within communication range. If additional packets are required to decode the file, these are downloaded from the MBS. We formulate an optimization problem that is efficiently solved numerically, providing TTL caching policies minimizing the overall network load. We demonstrate that distributed coded caching using TTL caching policies can offer significant reductions in terms of network load when request arrivals are bursty. We show how the distributed coded caching problem utilizing TTL caching policies can be analyzed as a specific single cache, convex optimization problem. Our problem encompasses static caching and the single cache as special cases. We prove that, interestingly, static caching is optimal under a Poisson request process, and that for a single cache the optimization problem has a surprisingly simple solution.
\end{abstract}


\section{Introduction}
Caching has attracted a significant amount of attention in the last few years as a promising technology to alleviate the load on backhaul links \cite{Boccardi2014}. Content may be cached in a distributed fashion across small base stations (SBSs) such that users can download requested content directly from them. For distributed caching, the use of erasure correcting codes (ECCs) has been shown to reduce the download delay as well as the network load \cite{Shanmugam2013, Bioglio2015}. Content may also be cached directly in mobile devices such that users can download content from neighboring devices using device-to-device (D2D) communication. Similar to the SBS caching case, the use of ECCs has been demonstrated to reduce the network load also for this scenario \cite{Pedersen2016, Pedersen2019, Wang2017}. Caching furthermore facilitates index-coded broadcasts to multiple users requesting different content, which has been shown to drastically reduce the amount of data that has to be transmitted over the SBS-to-device downlink \cite{Maddah-Ali2014}. ECCs have also been used in conjunction with index coding to ensure data availability at broadcasting devices in D2D caching networks \cite{Ji2016} as well as to provide index-coding broadcast opportunities to subsets of users \cite{Wei2017, Reisizadeh2018}. All these works consider the cached content to be static for a period of time (e.g., a day) according to a given file popularity distribution.

Dynamic cache eviction policies, e.g., first-in-first-out (FIFO), least-recently-used (LRU), least-frequently-used (LFU), and random (RND), may be beneficial to use when the file library or file popularity profile is dynamic, or when users request content according to a renewal process \cite{Gelenbe1973}. Due to the complexity in analyzing such policies, timer-based policies that are significantly more tractable have been suggested. One such policy is time-to-live (TTL) where a request for a particular piece of content triggers it to be cached and then evicted after the expiration of a timer. The TTL policy has been shown to yield similar performance to FIFO, LRU, LFU, and RND policies in \cite{Che2002, Fricker2012, Bianchi2013, Dehghan2019}. Goseling and Simeone extended the TTL policy to cache fractions of files, referred to as fractional TTL (FTTL), and showed that this can improve performance under a renewal request process \cite{Goseling2019}. Decreasing the fraction of a file that is cached over time, termed soft TTL (STTL), can further improve the performance. Optimal STTL caching policies are obtained through a convex optimization problem \cite{Goseling2019}. All previous works on TTL policies assume either a single cache or a number of caches, e.g., structured into lines or hierarchies, where users access a single cache. For these scenarios, coded caching does not bring any benefits. However, if users can access several caches, the use of ECCs can be beneficial. Hence, merging distributed coded caching with the TTL schemes in \cite{Goseling2019}, which have both independently been shown to bring performance improvements, is an intriguing prospect.

In this paper, we generalize the TTL policies  in \cite{Goseling2019} to a distributed coded caching scenario. Specifically, we consider the scenario where content is encoded using a maximum distance separable (MDS) code and cached in a distributed fashion across several SBSs. Coded content is evicted from the caches in accordance with the TTL policies in \cite{Goseling2019}. Users requesting a particular piece of content download coded packets from SBSs within communication range and, if necessary, download additional packets from a macro base station (MBS). The main contributions are summarized below.

\subsection{Contributions}
We generalize the TTL, FTTL, and STTL caching policies in \cite{Goseling2019} to a scenario where coded packets are cached in a distributed fashion across several SBSs. Specifically, we adopt the maximization problem in \cite{Goseling2019}, providing optimal caching policies for a single cache and an increasing and concave utility function. We modify the objective function of the problem in \cite{Goseling2019} to yield a network load minimization problem, where the network load is defined as a sum of data rates over various network links, weighted by a cost representing, e.g., transmission delay or energy consumption of transmitting data over these links. We then rewrite the optimization problem as a mixed integer linear program (MILP) that is efficiently solved numerically. We furthermore prove that the distributed coded caching problem can equivalently be analyzed as a single cache problem with a specific decreasing and convex cost function. This is an important result because it shows that such a function, previously studied for the single cache case due to its analytical tractability \cite{Goseling2019}, arises naturally in a distributed caching scenario. For SBSs deployed according to a Poisson point process \cite[Ch.~2.3]{Chiu2013}, we derive the cost function explicitly. We analyze two important special cases of the network load minimization problem. In particular, we show that our problem has the static coded caching problem where content is never updated (considered in, e.g., \cite{Shanmugam2013, Bioglio2015}), as a special case. We furthermore prove that static coded caching is optimal under the assumption of a Poisson request process. Moreover, for the special case of users accessing a single cache, we prove that the STTL problem is a fractional knapsack problem with a greedy optimal solution. The performance of TTL, FTTL, and STTL, in terms of network load, is evaluated for a renewal process, specifically when the times between requests follow a Weibull distribution, which has been shown to accurately model requests for, e.g., educational media \cite{Costa2004}. We show that distributed coded caching using TTL caching policies can offer significant reductions in network load, especially for bursty renewal request processes.

\subsection{Related Work}
Distributed caching of coded content utilizing TTL cache eviction policies was also investigated in \cite{Chen2019}. Compared to the problem studied in this paper, the work in \cite{Chen2019} is significantly different in a number of ways. Specifically, we consider an STTL policy with optimized TTL timers under a renewal request process, which was not considered in \cite{Chen2019}. Furthermore, a dynamic library of files with location-dependent popularity is considered in \cite{Chen2019}, which is typically considered to be more general than a static file library and is not in the scope of our work. However, it is reasonable to consider scenarios where the file library remains fixed for a considerable amount of time, e.g., a day, and focus on an area with homogeneous file popularity.

\section{System Model}
\label{sec:model}
\begin{figure}[!t]
	\centering{
		\includegraphics[width=\columnwidth]{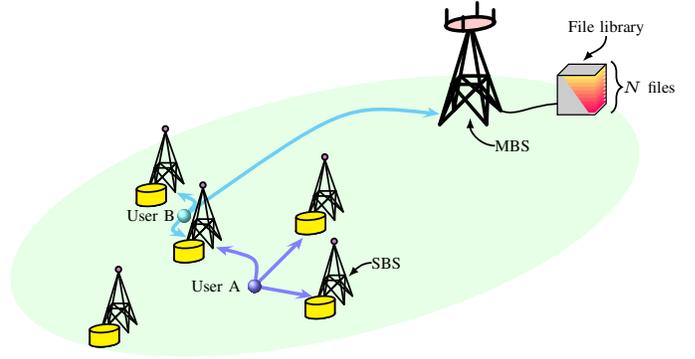}
		\caption{The considered network with an MBS with access to a file library of $N$ files, a number of SBSs caching coded fractions of files, and users A and B. In the example, user A can decode the requested file by downloading coded packets from the SBSs within communication range, whereas user B has to also download some packets from the MBS.}
		\label{fig:model}
		\vspace{-1em}
	}
\end{figure}

We consider an area served by an MBS that always has access to a file library of $N$ files, where file $\file = 1, 2, \ldots, N$ has size $s_i$. Mobile users request files from the library according to independent renewal processes. Specifically, we denote the independent and identically distributed times between requests for file $\file$ by $X_i$, the cumulative distribution function (CDF) of $X_i$ by
$$
	\cdf{X_i}{t} \triangleq \Pr(X_i \le t),
$$
and the request rate of file $\file$ by
$$
	\omega_i \triangleq \E[X_i]^{-1}.
$$
We let $p_i = \omega_i/\omega$, for some aggregate request rate in the area, $\omega = \sum_{i=1}^N \omega_i$. For a Poisson request process, i.e., exponentially distributed $X_i$, $p_i$ can be interpreted as the probability that file $\file$ is requested. The request rates $\omega_i$ are assumed to be constant over a sufficiently long period of time, e.g., not changing during the course of one day. For such scenarios, file popularity predictions and content allocation optimization can be carried out during periods of low network traffic, e.g., during night time. $\numsbs$ SBSs are deployed in the area and each SBS has a cache with storage capacity $C$. We assume that a user can download content from an SBS if it is within a range $\rsbs$ and we denote by $\gamma_b$ the probability that a user is within range of $b$ SBSs at any given time. The model considered in this paper is illustrated in Fig.~\ref{fig:model}.

\subsection{Caching Policy}
\label{sec:policy}
Each file $\file$ of size $s_i$ is partitioned into $k_i$ packets, each of size $s_i/k_i$. The packets are encoded into $n_i$ coded packets (also of size $s_i/k_i$) using an $(n_i, k_i)$ MDS code of rate $k_i/n_i$. We defer the description of how to select code parameters $n_i$ and $k_i$ until later in this subsection. For analytical tractability, we assume that all SBSs cache the same amount of each file at all times, i.e., the same number of distinct coded packets for a given file. In this respect, the caches are \emph{synchronized}. With slight abuse of notation, we let $m_i(t) \le k_i$ denote the number of distinct coded packets of file $\file$ cached at each SBS at time $t$, where $t$ is the time since the last request for file $\file$. We will use this interpretation of $t$ throughout the paper. The amount of file $\file$ cached by each SBS at time $t$ is hence $m_i(t) s_i/k_i$. We normalize by the file size $s_i$ and let
$$
	\mu_i(t) = m_i(t)/k_i
$$
denote the fraction of file $\file$ cached at time $t$. Similar to \cite{Goseling2019}, we refer to $\mu_i(t)$ as the \emph{caching policy}.

We adopt an STTL cache eviction policy, shown to increase the amount of content that can be downloaded from a single cache under a renewal request process in \cite{Goseling2019}. Hence, coded packets of file $\file$ may be evicted from the caches at periodic times with period $T$ after the last request for file $\file$. We allow $K$ potential updates within a total time equal to $KT$, which we refer to as the \emph{update window} length. For $K=0$, the caches are never updated. This corresponds to \emph{static caching}, which is the type of caching considered in a big part of the literature \cite{Shanmugam2013, Bioglio2015, Pedersen2016, Pedersen2019, Wang2017, Maddah-Ali2014}. The caching policies $\mu_i(t)$ are decreasing functions of $t$ given by \cite{Goseling2019}
\begin{equation}
	\mu_i(t) = \begin{cases}
		\mu_{i,0}, & \text{if}~t < T,\\
		\mu_{i,j}, & \text{if}~jT \le t < (j+1)T,~j = 1, 2, \ldots, K-1,\\
		\mu_{i,K}, & \text{if}~t \ge KT,
	\end{cases}
	\label{eq:mu}
\end{equation}
where
$$
	1 \ge \mu_{i,0} \ge \mu_{i,1} \ge \cdots \ge \mu_{i,K} \ge 0.
$$
To derive code parameters $k_i$ and $n_i$ from caching policy $\mu_{i,j}$ for any $i$, we first quantize $\mu_{i, j}^{-1}$, for all $j$, to rational numbers to the desired degree of precision. We then obtain $k_i$ as the least common multiple of $\mu_{i,j}^{-1}$, i.e., the smallest integer multiple of $\mu_{i,j}^{-1}$, for all $j$ such that $\mu_{i,j} > 0$, and $n_i = B k_i \mu_{i,0}$. After a request for file $i$, the $n_i$ coded packets are allocated to the caches, where each SBS caches $k_i \mu_{i,0}$ distinct coded packets. Subsequently, coded packets are evicted from the caches in accordance with the caching policy in \eqref{eq:mu}. The following small example illustrates how to obtain code parameters $k_i$ and $n_i$ from an STTL caching policy and how to manage cached content over time.
\begin{example}
	Consider a small network with three SBSs ($B=3$) and the STTL caching policy
	$$
		\mu_{i,j} = \begin{cases}
			1, & \text{if}~j = 0,\\
			2/3, & \text{if}~j = 1, 2, \ldots, 4,\\
			1/3, & \text{if}~j = 5,\\
			0, & \text{if}~j = K = 6,
		\end{cases}
	$$
	for some file $i$. The least common multiple of $1^{-1}$, $(2/3)^{-1}$, and $(1/3)^{-1}$ is $3$ and, hence, $k_i = 3$. It follows that $n_i = B k_i \mu_{i,0} = 9$. After a request for file $i$, each SBS caches $k_i \mu_{i,0} = 3$ distinct coded packets of file $i$. If there is no request for file $i$ until time $t=T$, one coded packet is evicted from the cache of each SBS, i.e., $k_i \mu_{i,1} = 2$ coded packets are cached at each SBS, etc..
\end{example}
We remark that, for MDS codes, the required underlying field size, as well as the encoding and decoding complexity, grow with $n_i$. If $n_i$ is large, rateless codes with significantly lower complexity can be used \cite{Luby2002}. Such codes have been shown to have close to MDS code performance in distributed caching scenarios \cite{Recayte2018}. In this work, we will assume that $\mu_{i,j} \in \R$ for simplicity. We refer to $f = 1/T$ as the update frequency and remark that static caching ($K=0$) corresponds to $f=0$.

\subsection{Content Download}
\label{sec:down}
For an MDS code, any $k_i$ coded packets of file $\file$ suffice to decode the file. All files requested by users can be decoded by downloading packets available at SBSs within communication range and, if necessary, retrieving additional packets from the MBS. Specifically, a user requesting file $\file$ at time $t$ downloads $m_i(t)$ packets from the $b$ SBSs within communication range. If $bm_i(t) \ge k_i$, the user can decode the file. If $b m_i(t) < k_i$, the additional $k_i-bm_i(t)$ coded packets required to decode the file are downloaded from the MBS. Consequently, the fraction of file $\file$ downloaded from SBSs can be expressed as
\begin{equation}
	\min\{1, b\mu_i(t)\}
	\label{eq:sbsfrac}
\end{equation}
and the fraction of file $\file$ downloaded from the MBS as
\begin{equation}
	\max\{0, 1-b\mu_i(t)\}.
	\label{eq:mbsfrac}
\end{equation}
We assume that downloading one bit of data from the MBS and the SBSs comes at a cost $\mbscost$ and $\sbscost$ per bit, respectively. The cost represents, e.g., the transmission delay or energy consumption of transmitting one bit. Finally, the cost to send data to the caches, referred to as the cache update cost, is denoted by $\updatecost$.

\section{Preliminaries}\label{sec:prel}
The caching policies in \eqref{eq:mu} correspond to STTL \cite{Goseling2019}. FTTL policies are obtained as a special case of \eqref{eq:mu}, where the same fraction $\nu_i$ of file $\file$ is cached for a time $LT$, defined by an integer $0 \le L \le K$, i.e., $\mu_{i,0} = \mu_{i,1} = \ldots = \mu_{i,L} = \nu_i$ and $\mu_{i,L+1} = \mu_{i,L+2} = \ldots = \mu_{i,K} = 0$ \cite{Goseling2019}. Furthermore, letting $\nu_i = 1$ we obtain TTL caching policies. In \cite{Goseling2019}, the caching problem is framed as a utility maximization problem where a strictly concave and increasing utility function $g_i(\mu)$ measures the utility resulting from caching a fraction $\mu$ when file $\file$ is requested. The choice of letting $g_i(\mu)$ be a strictly concave and increasing function of $\mu$ is due to analytical tractability. In practice, a linear utility function is more reasonable \cite{Neglia2018}.

For the case of a single cache, the sum utility maximization solved in \cite{Goseling2019} is
\begin{align}
	\underset{\substack{\mu_{i,j}, \nu_i \in \R\\ \beta_{i,j} \in \{0,1\}}}{\text{maximize}}~ & \sum_{i=1}^N \omega_i \sum_{j=0}^K g_i(\mu_{i,j}) F_{i,j}, \label{obj:goseling}\\
	\text{subject to}~ & \sum_{i=1}^N \omega_i s_i \sum_{j=0}^K \mu_{i,j} A_{i,j} \le C, \label{cnstr:cache}\\
		& 1 \ge \mu_{i,0} \ge \mu_{i,1} \ge \cdots \ge \mu_{i,K} \ge 0, \label{cnstr:mu}\\
		& 0 \le \nu_i \le 1, \label{cnstr:nu}\\
		& -\beta_{i,j} \le \mu_{i,j}\le \beta_{i,j}, \label{cnstr:beta1}\\
		& \beta_{i,j}-1 \le \mu_{i,j}-\nu_i \le 1-\beta_{i,j}, \label{cnstr:beta2}
\end{align}
where \eqref{cnstr:cache} is a long-term average cache capacity constraint \cite[Lem.~1]{Goseling2019},
\begin{equation}\label{eq:F}
	F_{i,j} = \begin{cases}
		\cdf{X_i}{(j+1)T}-\cdf{X_i}{jT}, & \text{if}~j = 0,\ldots, K-1,\\
		1-\cdf{X_i}{KT}, & \text{if}~j = K
	\end{cases}
\end{equation}
is the probability that file $i$ is requested in time-slot $j$, and
\begin{equation}\label{eq:A}
	A_{i,j} = \begin{cases}
		\displaystyle{\int}_{jT}^{(j+1)T} 1-\cdf{X_i}{t} \d t, & \text{if}~j = 0,\ldots, K-1,\\[1em]
		\displaystyle{\int}_{KT}^\infty 1-\cdf{X_i}{t} \d t, & \text{if}~j = K.
	\end{cases}
\end{equation}
The ratio $F_{i,j}/A_{i,j} \approx h_i(jT)$, where $h_i(\cdot)$ is the hazard function of the request process, represents the probability to observe a request given the time since the last request, and the approximation follows by considering the continuous limit $T \to 0$ \cite{Goseling2019}. For the remainder of this paper, $h_i(jT)$ is assumed to be decreasing in $j$. The solution to \eqref{obj:goseling}--\eqref{cnstr:beta2} provides optimal FTTL caching policies \cite{Goseling2019}. Optimal TTL caching policies are achieved by letting $\nu_i = 1$ and removing the constraint \eqref{cnstr:nu}, while STTL policies are achieved by removing the constraints \eqref{cnstr:nu}--\eqref{cnstr:beta2} \cite{Goseling2019}.

\section{Distributed Coded TTL Caching}
\label{sec:analysis}
In this section, we formulate the average rate at which data is sent through the network described in Section~\ref{sec:model} and an optimization problem to minimize the network load for coded TTL caching. In particular, we generalize the optimization problem \eqref{obj:goseling}--\eqref{cnstr:beta2} to a distributed coded caching scenario, utilizing TTL caching policies. We propose an equivalent, more tractable formulation of the optimization problem that is efficiently solved numerically. The average rate at which data is downloaded from the SBSs and the MBS is denoted by $\ratesbs$ and $\ratembs$, respectively. We choose the utility function $g_i(\mu_i(t)) = s_i \min\{1, b \mu_i(t)\}$, representing the amount of data that a user requesting file $i$ at time $t$ can download from the $b$ SBSs within communication range (see \eqref{eq:sbsfrac}). Using \eqref{eq:mu} and also averaging over the number of SBSs within range of a user requesting a particular content in \eqref{obj:goseling}, we obtain
\begin{equation}
	\ratesbs = \sum_{b=0}^{\numsbs} \gamma_b \sum_{i=1}^N \omega_i s_i \sum_{j=0}^K \min\{1, b\mu_{i,j}\} F_{i,j}.
	\label{eq:ratesbs}
\end{equation}
Similarly, substituting \eqref{eq:mu} in \eqref{eq:mbsfrac}, the MBS download rate is
\begin{equation}\label{eq:ratembsbeta}
	\ratembs = \sum_{b=0}^{\numsbs} \gamma_b \sum_{i=1}^N \omega_i s_i \sum_{j=0}^K \max\{0, 1-b\mu_{i,j}\} F_{i,j}.
\end{equation}
Note that $\max\{0, 1-b\mu_{i,j}\} = 1-\min\{1, b\mu_{i,j}\}$ and that, using \eqref{eq:F} together with $\cdf{X_i}{0} = 0$,
\begin{equation}\label{eq:sumF}
	\sum_{j=0}^K F_{i,j} = \cdf{X_i}{KT}-\cdf{X_i}{0}+1-\cdf{X_i}{KT} = 1.
\end{equation}
Hence, we may rewrite \eqref{eq:ratembsbeta} as
\begin{align}
	\ratembs & = \sum_{b=0}^{\numsbs} \gamma_b \sum_{i=1}^N \omega_i s_i \sum_{j=0}^K (1-\min\{1, b\mu_{i,j}\}) F_{i,j}\nonumber\\
		& = \sum_{i=1}^N \omega_i s_i - \ratesbs,
		\label{eq:ratembs}
\end{align}
i.e., all requested content not downloaded from SBSs is downloaded from the MBS. The average data rate at which the SBSs caches are updated, denoted by $\ratecache$, is
$$
	\ratecache = \numsbs \sum_{i=1}^N \omega_i s_i \sum_{j=0}^K (\mu_{i,0}-\mu_{i,j}) F_{i,j}.
$$
The above expression assumes that all caches are updated at each request in the area, which we refer to as \emph{synchronous} updates. This simplification is due to analytical tractability. Obtaining optimal caching policies under \emph{asynchronous} updates appears to be a formidable task. In Section~\ref{sec:results}, we nonetheless simulate caching policies that are optimal under synchronous updates for an asynchronous cache updating scenario.

We define the \emph{network load} as
\begin{equation}
	W = \mbscost\ratembs+\sbscost\ratesbs+\updatecost\ratecache,
	\label{eq:W}
\end{equation}
where
\begin{equation}
	\mbscost\ratembs+\sbscost\ratesbs = \mbscost \sum_{i=1}^N \omega_i s_i - (\mbscost-\sbscost) \ratesbs
	\label{eq:sumrate}
\end{equation}
using \eqref{eq:ratembs}. We want to minimize the network load over the caching policies $\mu_i(t)$ under the constraints \eqref{cnstr:cache}--\eqref{cnstr:beta2}. The first term in \eqref{eq:sumrate} is independent of $\mu_i(t)$. Hence, minimizing \eqref{eq:W} is equivalent to minimizing
$$
	\updatecost\ratecache - (\mbscost-\sbscost) \ratesbs.
$$
Thus, minimizing the network load corresponds to the optimization problem
\begin{align}
	\underset{\substack{\mu_{i,j}, \nu_i \in \R\\ \beta_{i,j} \in \{0,1\}}}{\text{minimize}}~ & \updatecost\ratecache - (\mbscost-\sbscost) \ratesbs, \label{obj:min}\\
	\text{subject to}~ & \text{\eqref{cnstr:cache}--\eqref{cnstr:beta2}}. \nonumber
\end{align}
We remind that optimal code parameters $n_i$ and $k_i$ are readily obtained from the caching policies $\mu_{i,j}$ minimizing the network load (see Section~\ref{sec:policy}). Consider briefly the case of zero cache update cost, i.e., $\updatecost = 0$. For $\mbscost > \sbscost$, we see that \eqref{obj:min} represents a maximization of the SBS download rate $\ratesbs$ and for $\mbscost \le \sbscost$, \eqref{obj:min} has a trivial solution $\mu_{i,j} = 0$, i.e., caching at the SBSs is turned off ($\ratesbs = 0$) and all data is fetched from the MBS, for which
$$
	W = \mbscost \sum_{i=1}^N \omega_i s_i
$$
using \eqref{eq:W} and \eqref{eq:sumrate}.

Next, we reformulate the optimization problem \eqref{obj:min} in a way that is more tractable. Using the epigraph formulation \cite[Ch.~3.1.7]{Boyd2009}, we introduce the auxiliary optimization variables $\xi_{b,i,j} \in \R$ and the constraints
\begin{align}
	\xi_{b,i,j} & \le 1, \label{cnstr:xi1}\\
	\xi_{b,i,j} & \le b \mu_{i,j}. \label{cnstr:xi2}
\end{align}
Expressing the SBS download rate in \eqref{eq:ratesbs} as 
\begin{equation}\label{eq:ratesbstilde}
	\ratetilde = \sum_{b=0}^B \gamma_b \sum_{i=1}^N \omega_i s_i \sum_{j=0}^K \xi_{b,i,j} F_{i,j},
\end{equation}
with the notation $\ratetilde$ to emphasize that it corresponds to the download rate of the epigraph formulation, we formulate the MILP
\begin{align}
	\underset{\substack{\mu_{i,j}, \nu_i, \xi_{b,i,j} \in \R\\ \beta_{i,j} \in \{0,1\}}}{\text{minimize}}~ & \updatecost\ratecache - (\mbscost-\sbscost) \ratetilde, \label{obj:epi}\\
	\text{subject to}~ & \text{\eqref{cnstr:cache}--\eqref{cnstr:beta2}, \eqref{cnstr:xi1}, \eqref{cnstr:xi2}},\nonumber
\end{align}
which is equivalent to \eqref{obj:min} and efficiently solved using, e.g., Gurobi \cite{Gurobi2018}. The MILP \eqref{obj:epi} provides optimal FTTL caching policies for the distributed coded caching scenario. Optimal coded TTL policies are achieved by letting $\nu_i = 1$ and removing the constraint \eqref{cnstr:nu}, while coded STTL policies are attained by removing the constraints \eqref{cnstr:nu}--\eqref{cnstr:beta2}. Note that the STTL optimization problem is a linear program. We observe that the optimal STTL caching policies $\mu_{i,j}$ are almost exclusively rational numbers which removes the quantization step when obtaining code parameters $k_i$ and $n_i$, as explained in Section~\ref{sec:policy}.

\subsection{Analysis as Single Cache TTL}
\label{sec:transform}
In this subsection, we will show that the distributed coded caching problem using TTL caching policies in \eqref{obj:min} can equivalently be analyzed as a single cache TTL problem using a particular decreasing and convex cost function. We also show how our distributed caching problem maps to the sum utility maximization \eqref{obj:goseling} for the single cache case. Let the random variable $Y$ denote the number of SBSs within range of a user, with $\Pr(Y = b) = \gamma_b$, $b = 0, 1, \ldots, \numsbs$. Changing the order of summation in \eqref{eq:ratesbs} yields
\begin{align}
	\ratesbs & = \sum_{i=1}^N \omega_i s_i \sum_{j=0}^K F_{i,j} \E[\min\{1, \mu_{i,j} Y\}].\label{eq:sumb}
\end{align}
Regarding the expectation in \eqref{eq:sumb}, we will need the following lemma in subsequent theorems.
\begin{lemma}\label{lem:exp}
	For a nonnegative random variable $Y$ and $\mu \ge 0$,
	$$
		\E[\min\{1, \mu Y\}] = \int_0^1 1 - F_Y(z/\mu) \d z.
	$$
\end{lemma}
\begin{IEEEproof}
	See Appendix~\ref{prf:lemexp}.
\end{IEEEproof}

The following theorem gives some important properties of the expectation in \eqref{eq:sumb}, as a function of the caching policy $\mu_{i,j}$.
\begin{theorem}\label{th:exp}
	For a nonnegative random variable $Y$, the expectation $\E[\min\{1, \mu Y\}]$ is an increasing and concave function of $\mu \ge 0$.
\end{theorem}
\begin{IEEEproof}
	See Appendix~\ref{prf:exp}.
\end{IEEEproof}
The result of Theorem~\ref{th:exp} is interesting because it proves that \eqref{obj:min} is convex. Furthermore, it shows how the cost minimization \eqref{obj:min} with link costs $\mbscost = 1$, $\sbscost = 0$, and no cache update cost ($\updatecost = 0$), which corresponds to a distributed caching scenario, maps to the utility maximization \eqref{obj:goseling}, which assumes a single cache. The following theorem considers the important special case of SBSs distributed in an area according to a Poisson point process, in which case $Y$ corresponds to a Poisson random variable \cite[Ch.~2.3]{Chiu2013}.
\begin{theorem}\label{th:poissexp}
	For $Y\sim\textnormal{Poisson}(\lambda)$,
	\begin{equation}
	\label{eq:poissexp}
		\E[\min\{1, \mu Y\}] = 1+(\lambda\mu-1) Q(\ceil{1/\mu}, \lambda)-\frac{\e^{-\lambda} \lambda^{\ceil{1/\mu}} \mu}{\Gamma(\ceil{1/\mu})},
	\end{equation}
	where $Q(\cdot, \cdot)$ is the regularized Gamma function and $\Gamma(\cdot)$ is the Gamma function.
\end{theorem}
\begin{IEEEproof}
	See Appendix~\ref{prf:poissexp}.
\end{IEEEproof}

The expression \eqref{eq:poissexp} is an increasing and concave function of $\mu$, according to Theorem~\ref{th:exp}. Due to the ceiling function $\ceil{1/\mu}$ in \eqref{eq:poissexp}, we see that we should set $1/\mu \in \N$ in order to minimize \eqref{obj:min} while not wasting cache capacity resources (see \eqref{cnstr:cache}).

\section{Special Cases}
The distributed coded caching problem utilizing TTL caching policies \eqref{obj:min} has two interesting problems as special cases; static caching ($K = 0$), studied in \cite{Shanmugam2013, Bioglio2015, Pedersen2019, Wang2017} for MDS codes, and single cache ($\gamma_1 = 1$), investigated in \cite{Goseling2019}. In this section, we show the connection between our problem and the special case of static caching, which we prove is optimal under a Poisson request process, and the special case of a single cache, which we prove has a particularly simple optimal solution.

\subsection{Static Coded Caching}
Before showing that \eqref{obj:min} includes static caching as a special case, we have the following theorem.
\begin{theorem}
\label{th:static}
	For a Poisson request process, static caching minimizes \eqref{obj:min}.
\end{theorem}
\begin{IEEEproof}
	See Appendix~\ref{prf:static}.
\end{IEEEproof}
Under static caching, FTTL and STTL are identical as only the updates distinguish the two caching policies. In the following, we assume that all files are of equal size, i.e., $s_i = s$, and study the nontrivial case $\mbscost > \sbscost$. For static caching ($K=0$), \eqref{eq:F} reduces to
$$
	F_{i,j} = F_{i,0} = 1 - \cdf{X_i}{0} = 1
$$
and the objective function \eqref{obj:min} becomes
\begin{align}
	-\ratesbs & = -\sum_{b=0}^{\numsbs} \gamma_b \sum_{i=1}^N \omega_i s_i \min\{1, b\mu_{i,0}\} F_{i,0}\nonumber\\
		& = -\omega s \sum_{b=0}^{\numsbs} \gamma_b \sum_{i=1}^N p_i \min\{1, b\mu_{i,0}\}.\label{obj:bioglio}
\end{align}
Also, \eqref{eq:A} is
$$
	A_{i,j} = A_{i,0} = \int_0^\infty 1-\cdf{X_i}{t} \d t = \E[X_i] = \omega_i^{-1}.
$$
Hence, the constraint \eqref{cnstr:cache} simplifies to
\begin{equation}
	\sum_{i=1}^N \omega_i s_i \sum_{j=0}^K \mu_{i,j} A_{i,j} = s \sum_{i=1}^N \mu_{i,0} \le C.
	\label{cnstr:bioglio}
\end{equation}
Using \eqref{obj:bioglio} in \eqref{obj:min}, under constraints \eqref{cnstr:mu} and \eqref{cnstr:bioglio}, the optimization problem \eqref{obj:min} is precisely the static caching problem considered in \cite{Bioglio2015}, apart from additive and multiplicative constants. Hence, the static caching problem explored in \cite{Bioglio2015} is a special case of \eqref{obj:min}.

\subsection{Single Cache TTL}
We proceed with the other interesting special case, i.e., the single cache problem. Letting $\gamma_1 = 1$ in \eqref{eq:ratesbs}, i.e., users access a single cache with probability 1, we see that the average rate at which data is downloaded from the SBSs is
\begin{align}
	\ratesbs & = \sum_{i=1}^N \omega_i s_i \sum_{j=0}^K \min\{1, \mu_{i,j}\} F_{i,j} \nonumber\\
		& = \sum_{i=1}^N \omega_i s_i \sum_{j=0}^K \mu_{i,j} F_{i,j}, \label{obj:minsingle}
\end{align}
since $\mu_{i,j} \le 1$ using \eqref{cnstr:mu}. For the nontrivial case of $\mbscost > \sbscost$, and update cost $\updatecost = 0$, the minimization problem \eqref{obj:min} is equivalent to a maximization problem of the objective function \eqref{obj:minsingle}. In particular, the STTL problem has a surprisingly simple solution given by the following theorem.
\begin{theorem}
\label{th:fracknap}
	For users accessing a single cache, link costs $\mbscost > \sbscost$ and $\updatecost = 0$, the STTL optimization problem, i.e., maximizing \eqref{obj:minsingle} under constraints \eqref{cnstr:cache} and \eqref{cnstr:mu}, is a fractional knapsack problem with a greedy optimal solution equivalent to FTTL and TTL.
\end{theorem}
\begin{IEEEproof}
	See Appendix~\ref{prf:fracknap}.
\end{IEEEproof}
A similar result was proved in \cite{Goseling2019} for a single file, i.e., $N=1$.

%

\begin{figure}[!t]
	\centering
%
%
%
%
%
	\includegraphics[width=\columnwidth]{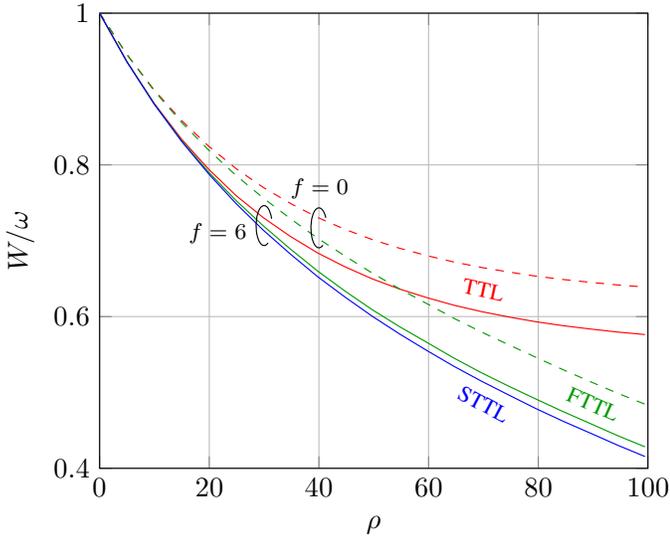}
	\caption{The fraction of data downloaded from the MBS as a function of the SBS density $\rho$.}
	\label{fig:density}
	\vspace{-2ex}
\end{figure}

\section{Numerical Results}
\label{sec:results}
In the following, we will assume that the times between requests are distributed according to the Weibull distribution, which has been shown to accurately estimate inter-request times of, e.g., educational media \cite{Costa2004}, i.e., $X_i \sim \text{Weibull}(a_i, b_i)$, where $a_i$, $0 < a_i \le 1$, and $b_i$ are the shape and scale parameters of the distribution, respectively. For simplicity, we let $a_i = a$ for all $i$ to evaluate the network load for a class of files requested with the same Weibull shape statistics. However, we remark that our optimization framework can handle general $a_i$. The CDF of the Weibull distribution is
$$
	\cdf{X_i}{t} = 1 - \exp\left[-\left(\frac{t}{b_i}\right)^{a}\right].
$$
Also,
$$
	\omega_i^{-1} = \E[X_i] = b_i \Gamma(1+a^{-1}),
$$
which implies
$$
	b_i = \frac{1}{\omega_i \Gamma(1+a^{-1})}.
$$
We assume the aggregate request rate per hour $\omega = 100$ and
$$
	p_i = \frac{1/i^\alpha}{\sum_{\ell=1}^N 1/\ell^\alpha},
$$
which is the Zipf probability mass function with parameter $\alpha \ge 0$. We remind that $p_i$ has the interpretation of file popularity under a Poisson request process.

The area is defined by the communication range of the MBS, which is denoted by $\rmbs$ and assumed to be $\rmbs = 800$ meters (m), i.e., the considered area is $\pi \rmbs^2 \approx 2$ square kilometers. The SBSs are deployed in the area according to a Poisson point process. Let $\rho$ be the density of SBSs per square kilometer (km$^{-2}$), i.e., $\rho = \numsbs/(\pi \rmbs^2)$. The probability that a user is within range of $b$ SBSs is \cite[Ch.~2.3]{Chiu2013}
$$
	\gamma_b = \e^{-\lambda} \frac{\lambda^b}{b!},
$$
where $\lambda = \rho \pi \rsbs^2 = \numsbs (\rsbs/\rmbs)^2$. Unless stated otherwise, we will assume the following setup for the remainder of this section. The library holds $N = 100$ files, each of normalized size $s_i = 1$. We set the Weibull shape $a = 0.6$, which describes a quite bursty request process and is within the range specified in \cite{Costa2004}. Also, we set $\alpha = 0.7$, which has been shown to accurately capture the popularity of Youtube videos \cite{Cheng2008}. We assume that there are $B = 100$ SBSs in the area, corresponding to a density $\rho \approx 50$ km$^{-2}$ and that users can download content from SBSs within a range of $\rsbs = 100$ m. Each SBS has the capacity to cache $C = 10$ files or $10\%$ of the file library. We assume the link costs $\sbscost = 0$ and $\mbscost = 1$. Furthermore, we assume that $\updatecost \ll \mbscost$, which is a reasonable assumption since data can be transmitted to caches over high capacity fiber-optical or highly directional wireless backhaul links, while the MBS serves a large number of users over potentially large distances. Finally, we consider an update window length of $K/f = 1$ hour and update frequencies $f = 6$ per hour.

\begin{figure}[!t]
	\centering
	\includegraphics[width=\columnwidth]{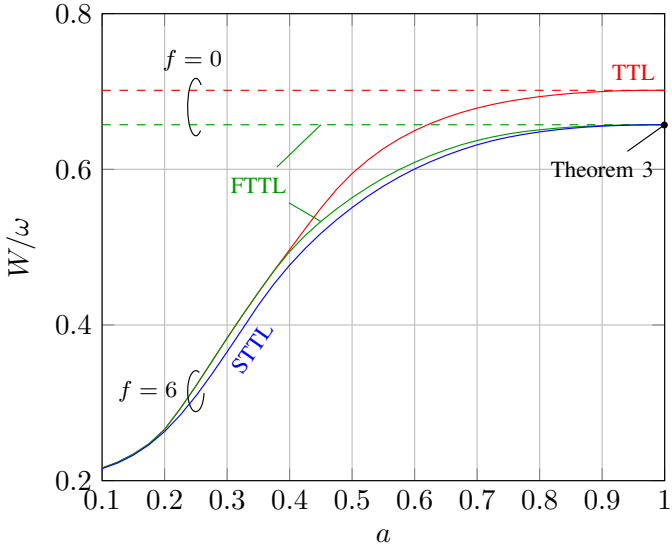}
	\caption{The fraction of data downloaded from the MBS as a function of the Weibull shape $a$ for an update frequency $f=6$.}
	\label{fig:shape}
	\vspace{-2ex}
\end{figure}

We obtain optimal TTL, FTTL, and STTL caching policies by solving \eqref{obj:epi} and plot the network load \eqref{eq:W} normalized by the aggregate request rate $\omega$. For $\updatecost = 0$, the network load is interpreted as the fraction of data downloaded from the MBS. Fig.~\ref{fig:density} shows this fraction as a function of the SBS density $\rho$ for no cache updates, i.e., $f = 0$ implying $\mu_{i,j} = \mu_{i,0}$, and cache update frequency per hour $f = 6$. The network load using FTTL and STTL overlap for $f = 0$, which is expected since only the cache updates distinguish the two policies. We also see that there is a reduction in network load when choosing the FTTL or STTL caching policies over the TTL policy and that the network load decreases with increasing SBS density. The reason for the performance loss when using the TTL caching policy is that users within range of $b>1$ SBSs will download superfluous data, which correspond to a wasteful use of cache memory resources. The gain for the static caching scenario ($f = 0$) was observed already in \cite{Bioglio2015}. Finally, we observe that, for $f=6$, there is only a small reduction in network load for STTL as compared to FTTL, but the load reduction is increasing for increasing $\rho$.

Fig.~\ref{fig:shape} shows the fraction of data downloaded from the MBS versus the Weibull shape $a$, for update frequency per hour $f=6$, and no cache update cost ($\updatecost=0$). We also include curves for static caching ($f=0$), which do not depend on $a$ as is shown in \eqref{obj:bioglio} and \eqref{cnstr:bioglio}, for comparison. For bursty request arrivals, i.e., small values of $a$, we see that the use of TTL caching policies reduces the fraction of data downloaded from the MBS significantly with respect to static caching. Furthermore, we observe that, for very small values of $a$, all TTL policies have similar performance. This is because, with high probability, the times between requests are less than the period $T$, i.e., $F_{i,0} \approx 1$ for all $i$, and TTL is an optimal caching policy. For $a=1$, corresponding to a Poisson request process, FTTL and STTL yield the same network load as proved in Theorem~\ref{th:static}, which is, however, lower than the network load using TTL. A similar effect was shown in \cite{Goseling2019} for the single cache case.

\begin{figure}[!t]
	\centering
	\includegraphics[width=\columnwidth]{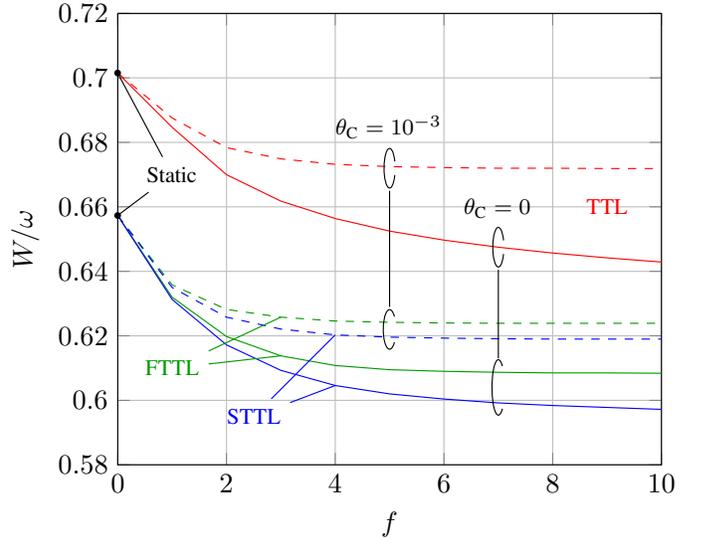}
	\caption{Normalized network load as a function of the cache update frequency $f$.}
	\label{fig:updatefreq}
	\vspace{-2ex}
\end{figure}

Fig.~\ref{fig:updatefreq} shows the normalized minimum network load as a function of the update frequency $f$ for the case of no cache update cost ($\updatecost = 0$) and $\updatecost = 10^{-3}$. For both cases, updating content on the caches is seen to be beneficial for all TTL policies. For example, using STTL and assuming $\updatecost = 0$, the reduction is roughly 10\% as compared to static caching. We observe that the decrease in network load for an increase in update frequency saturates for moderately large $f$. Hence, cache updates need not be very frequent to reap the benefits of the TTL, FTTL, and STTL caching policies. The sufficient update frequency of course depends on several parameter values, in particular, the Weibull shape $a$ is a key parameter when deciding update frequencies.

\begin{figure}[!t]
	\centering
	\includegraphics[width=\columnwidth]{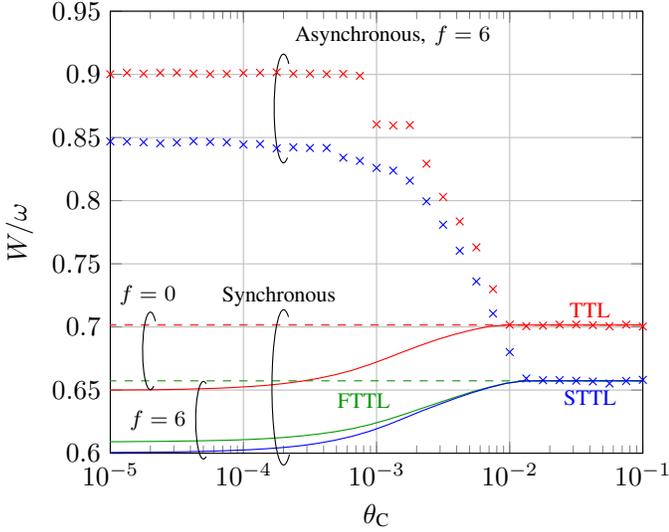}
	\caption{Normalized network load versus the update cost $\updatecost$ when using the various caching policies under synchronous and asynchronous cache updates.}
	\label{fig:updatecost}
	\vspace{-2ex}
\end{figure}

Finally, in Fig.~\ref{fig:updatecost}, the normalized minimum network load is plotted versus the cache update cost $\updatecost$ for an update frequency per hour $f=6$. The load for static caching ($f=0$) is also shown in the figure. As previously described, the network load when using FTTL and STTL is the same for static caching. It is interesting to note that all TTL policies revert to static caching for sufficiently large values of $\updatecost$. Also included in Fig.~\ref{fig:updatecost} is a simulation of the optimal (under synchronous cache updates) STTL and TTL caching policies for asynchronous cache updates, i.e., only the SBSs within range of a user placing a request update cached content. The considered caching policies do not exhibit a better performance under asynchronous updates, which is to be expected for two reasons. Firstly, since the file request process is homogenous over the considered area, the spatial average cached content is important and the same average cached content can be achieved by both synchronous and asynchronous cache updates. Secondly, the request rate within the communication range of an SBS is smaller than the request rate in the entire area, implying less content to be cached over time using asynchronous updates, i.e., the caches are underutilized.

\section{Conclusion}
We optimized time-to-live (TTL) caching policies with periodic eviction of coded content to minimize the overall network load for a scenario where content is encoded using a maximum distance separable (MDS) code and cached in a distributed fashion across small base stations. The proposed optimization problem is efficiently solved numerically. Interestingly, we show that the problem can equivalently be analyzed as a single cache optimization problem under a specific decreasing and convex cost function. For small base stations (SBSs) deployed according to a Poisson point process, we provide the cost function explicitly. The analyzed scenario encompasses static caching and single caching as important special cases. We proved that, interestingly, static caching is optimal under a Poisson request process. We also proved that the single cache problem has a simple greedy solution. We showed that TTL caching policies can offer substantial reductions in network load compared with static caching under a request renewal process, in particular when the request process is bursty. Conversely, for sufficiently large cache update cost, dynamic caching of MDS coded content is futile, i.e., static caching is optimal. Finally, although we consider a wireless network scenario, the results are general in the sense that they can be applied to any distributed caching scenario.

The updating of coded content cached at SBSs can be seen as content repair \cite{Pedersen2016}. Therefore, an interesting question is whether repair-efficient coding schemes, e.g., regenerating codes \cite{Dimakis2010}, can yield a lower network load.

\begin{appendices}
\section{Proof of Lemma~\ref{lem:exp}}
\label{prf:lemexp}
We represent $1$ as a random variable with degenerate distribution $\delta(z-1)$, where $\delta(\cdot)$ is the Dirac delta function, and let $Z = \min\{1, \mu Y\}$, for which the CDF of $Z$ is
\begin{align*}
	F_Z(z) & \triangleq \Pr(Z \le z)\\
		& = \Pr(\min\{1, \mu Y\} \le z)\\
		& = 1-\Pr(1>z, \mu Y > z)\\
		& = 1-H(1-z) (1-F_Y(z/\mu)),
\end{align*}
where $H(\cdot)$ is the heavyside function. The expected value of $Z$ is
\begin{align*}
	\E[Z] & = \int_0^\infty 1-F_Z(z) \d z\\
		& = \int_0^\infty H(1-z) (1-F_Y(z/\mu)) \d z\\
		& = \int_0^1 1- F_Y(z/\mu) \d z.
\end{align*}

\section{Proof of Theorem~\ref{th:exp}}
\label{prf:exp}
Since $F_Y(y)$ is an increasing function of $y$, $1-F_Y(z/\mu)$ is an increasing function of $\mu$, and
$$
	\int_0^1 1-F_Y(z/\mu) \d z
$$
is an increasing function of $\mu$. Using Lemma~\ref{lem:exp}, the expectation $\E[\min\{1, \mu Y\}]$ is an increasing function of $\mu$.
	
For $\mu_1 \ge 0$, $\mu_2 \ge 0$, $Y \ge 0$, and $0 \le \alpha \le 1$, the following inequalities hold,
\begin{align*}
	(1-\alpha) \mu_1 Y & \ge (1-\alpha) \min\{1, \mu_1 Y\},\\
	\alpha \mu_2 Y & \ge \alpha \min\{1, \mu_2 Y\}.
\end{align*}
Hence,
\begin{equation}
	((1-\alpha) \mu_1 + \alpha \mu_2) Y  \ge (1-\alpha) \min\{1, \mu_1 Y\} + \alpha \min\{1, \mu_2 Y\}.
	\label{eq:conv1}
\end{equation}
Similarly, using
\begin{align*}
	(1-\alpha) & \ge (1-\alpha) \min\{1, \mu_1 Y\},\\
	\alpha & \ge \alpha \min\{1, \mu_2 Y\},
\end{align*}
we have that
\begin{equation}
	1 = 1-\alpha+\alpha \ge (1-\alpha) \min\{1, \mu_1 Y\} + \alpha \min\{1, \mu_2 Y\}.
	\label{eq:conv2}
\end{equation}
Using \eqref{eq:conv1} and \eqref{eq:conv2}, we get
\begin{align*}
	& \min\{1, ((1-\alpha) \mu_1 + \alpha \mu_2) Y\}\\
	& \hspace{4em} \ge (1-\alpha) \min\{1, \mu_1 Y\} + \alpha \min\{1, \mu_2 Y\}.
\end{align*}
Taking the expectation of both sides yields
\begin{align*}
	& \E[\min\{1, ((1-\alpha) \mu_1 + \alpha \mu_2) Y\}]\\
	& \hspace{4em} \ge \E[(1-\alpha) \min\{1, \mu_1 Y\} + \alpha \min\{1, \mu_2 Y\}]\\
	& \hspace{4em} = (1-\alpha) \E[\min\{1, \mu_1 Y\}] + \alpha \E[\min\{1, \mu_2 Y\}],
\end{align*}
which concludes the proof.

\section{Proof of Theorem~\ref{th:poissexp}}
\label{prf:poissexp}
For a Poisson random variable $Y$ with rate $\lambda$, 
\begin{equation}\label{eq:ycdf}
	F_Y(y) = \sum_{i=0}^{\floor{y}} \frac{\lambda^i}{i!} \e^{-\lambda}.
\end{equation}
For a positive integer $x$, let $\Gamma(x)$ and $Q(x, \lambda)$ denote the Gamma function and the regularized Gamma function, i.e., $\Gamma(x) = (x-1)!$ and
\begin{equation}\label{eq:qrec}
	Q(x+1, \lambda) = \int_\lambda^\infty \frac{t^x \e^{-t}}{\Gamma(x+1)} \d t \overset{(a)}{=} \frac{\lambda^x \e^{-\lambda}}{\Gamma(x+1)} + Q(x, \lambda),
\end{equation}
respectively, where $(a)$ is obtained after integration by parts. Unfolding the recursion in \eqref{eq:qrec}, using $Q(1, \lambda) = \e^{-\lambda}$ yields
\begin{equation}\label{eq:qsum}
	Q(x+1, \lambda) = \sum_{i=0}^x \frac{\lambda^i}{\Gamma(i+1)} \e^{-\lambda} = \sum_{i=0}^x \frac{\lambda^i}{i!} \e^{-\lambda}
\end{equation}
and
\begin{equation}\label{eq:qcdf}
	Q(\floor{y}+1, \lambda) = F_Y(y),
\end{equation}
using \eqref{eq:ycdf}.

Using \eqref{eq:qcdf},
\begin{align}
	& \int_0^1 F_Y(y/\mu) \d y = \int_0^1 Q(\floor{y/\mu}+1, \lambda) \d y\nonumber\\
		& \hspace{2em} = \mu \sum_{i=1}^{\floor{1/\mu}} Q(i, \lambda) + (1-\floor{1/\mu}\mu) Q(\floor{1/\mu}+1, \lambda),\label{eq:cdfint}
\end{align}
where the integral is a summation due to the floor function in the argument of $Q(\cdot, \cdot)$. Applying the recursion \eqref{eq:qrec} repeatedly yields
\begin{align*}
	\sum_{i=1}^{\floor{1/\mu}} Q(i, \lambda) & = \floor{1/\mu} Q(\floor{1/\mu}+1, \lambda) - \sum_{i=1}^{\floor{1/\mu}} i \frac{\lambda^i \e^{-\lambda}}{\Gamma(i+1)}\\
		& = \floor{1/\mu} Q(\floor{1/\mu}+1, \lambda) - \lambda \sum_{j=0}^{\floor{1/\mu}-1} \frac{\lambda^{j} \e^{-\lambda}}{\Gamma(j+1)}\\
		& \overset{(b)}{=} \floor{1/\mu} Q(\floor{1/\mu}+1, \lambda) - \lambda Q(\floor{1/\mu}, \lambda),
\end{align*}
where we have used \eqref{eq:qsum} in $(b)$. Inserting this expression in \eqref{eq:cdfint}, one obtains
\begin{align}
	\int_0^1 F_Y(y/\mu) \d y & = Q(\floor{1/\mu}+1, \lambda) - \mu \lambda Q(\floor{1/\mu}, \lambda)\nonumber\\
		& \overset{(c)}{=} (1-\lambda\mu) Q(\ceil{1/\mu}, \lambda) + \frac{\lambda^{\ceil{1/\mu}} \mu \e^{-\lambda}}{\Gamma(\ceil{1/\mu})},\label{eq:cdfint2}
\end{align}
where we have used \eqref{eq:qrec} and
$$
	\ceil{1/\mu}-\floor{1/\mu} = \begin{cases}
		0, & \text{if}~1/\mu \in \mathbb{Z}\\
		1, & \text{if}~1/\mu \notin \mathbb{Z}
	\end{cases}
$$
in $(c)$. Combining \eqref{eq:cdfint2} with the result of Lemma~\ref{lem:exp} yields the desired result.

\section{Proof of Theorem~\ref{th:static}}
\label{prf:static}
For proving that static caching minimizes \eqref{obj:epi} (and hence \eqref{obj:min}) it is sufficient to show that it maximizes $\ratetilde$ (see \eqref{eq:ratesbstilde}), as for static caching $R_C = 0$. Maximizing \eqref{eq:ratesbstilde} under constraints \eqref{cnstr:cache}, \eqref{cnstr:mu}, \eqref{cnstr:xi1}, and \eqref{cnstr:xi2}, is equivalent to 
\begin{align}
	\underset{\mu_{i,j}, \xi_{b,i,j}, C_i \in \R}{\text{maximize}}~ & \sum_{b=0}^B \gamma_b \sum_{i=1}^N \omega_i s_i \sum_{j=0}^K \xi_{b,i,j} F_{i,j},\label{obj:sbs}\\
	\text{subject to}~ & s_i \sum_{j=0}^K \mu_{i,j} F_{i,j} \le C_i,\label{cnstr:cache1}\\
		& \sum_{i=1}^N C_i = C,\label{cnstr:cache2}\\
		& \mu_{i,j} - \mu_{i,j-1} \le 0,~\mu_{i,-1} = 1,\\
		& -\mu_{i,j} \le 0,\\
		& \xi_{b,i,j} \le 1,\\
		& \xi_{b,i,j} \le b \mu_{i,j},\label{cnstr:xi2sbs}
\end{align}
where $C_i$ may be regarded as the size of the cache partition reserved for file $i$, and we used the fact that for exponentially distributed inter-request times $F_{i,j}/A_{i,j} = \omega_i$ (from \eqref{eq:F} and \eqref{eq:A}), i.e., the hazard function is constant for a Poisson request process. For fixed $C_i$'s, the maximization problem is separable in $i$. Thus, we can consider the following optimization problem
\begin{align}
	\underset{\mu_{i,j}, \xi_{b,i,j} \in \R}{\text{maximize}}~ & \sum_{b=0}^B \gamma_b \sum_{j=0}^K \xi_{b,i,j} F_{j}, \label{obj:single}\\
	\text{subject to}~ & s_i \sum_{j=0}^K \mu_{i,j} F_{i,j} \le C_i, \label{cnstr:cachesingle}\\
		& \mu_{i,j} - \mu_{i,j-1} \le 0,~\mu_{i,-1} = 1, \label{cnstr:mu1single}\\
		& -\mu_{i,j} \le 0, \label{cnstr:mu2single}\\
		& \xi_{b,i,j} \le 1, \label{cnstr:xi1single}\\
		& \xi_{b,i,j} \le b \mu_{i,j}, \label{cnstr:xi2single}
\end{align}
for each file $i=1, \ldots, N$ separately. We can now prove the following lemma.
\begin{lemma}\label{lem:kkt}
	Static caching is an optimal solution to \eqref{obj:single}--\eqref{cnstr:xi2single}.
\end{lemma}
\begin{IEEEproof}
For ease of exposition, we drop the subindex $i$ in the proof. Introducing the dual variables $\lambda$, $\phi_j$, $\psi_j$, $\delta_{b,j}$, and $\epsilon_{b,j}$, the Karush-Kuhn-Tucker (KKT) conditions of \eqref{obj:single}--\eqref{cnstr:xi2single} are
\begin{align}
	-\gamma_b F_j + \delta_{b,j} + \epsilon_{b,j} & = 0,\\
	\lambda s F_j + \phi_j - \phi_{j+1} - \psi_j - \sum_{b=0}^B \epsilon_{b,j} b & = 0,~\phi_{K+1} = 0,\\
	\lambda \left( - C + s \sum_{j=0}^K \mu_j F_j \right) & = 0,\\
	\phi_j (\mu_j-\mu_{j-1}) & = 0,~\mu_{-1} = 1,\\
	\psi_j (-\mu_j) & = 0,\\
	\delta_{b,j} (\xi_{b,j}-1) & = 0,\\
	\epsilon_{b,j} (\xi_{b,j}-b\mu_j) & =  0,\\
	\lambda \ge 0, \phi_j \ge 0, \psi_j & \ge 0,\\
	\delta_{b,j} \ge 0, \epsilon_{b,j} & \ge 0,
\end{align}
and \eqref{cnstr:cachesingle}--\eqref{cnstr:xi2single}. Let
\begin{align*}
	\mu_j & = C/s,\\
	\xi_{b,j} & = \min\{1, b C/s\},
\end{align*}
which corresponds to static caching utilizing completely the given cache partition, and largest possible values of the variables $\xi_{b,j}$. Furthermore, let
\begin{align}
	\phi_j & = 0,\\
	\psi_j & = 0,\\
	\delta_{b,j} & = \begin{cases}
		0, & \text{if}~b \le s/C\\
		\gamma_b F_j, & \text{if}~b > s/C
	\end{cases},\\
	\epsilon_{b,j} & = \begin{cases}
		\gamma_b F_j, & \text{if}~b \le s/C\\
		0, & \text{if}~b > s/C
	\end{cases},\label{eq:epsilon}\\
	\lambda & = \frac{1}{s} \sum_{j=0}^K \sum_{b=0}^B \epsilon_{b,j} b \overset{(a)}{=} \frac{1}{s} \sum_{b=0}^{\floor{s/C}} \gamma_b b,
\end{align}
where we have used \eqref{eq:epsilon} and \eqref{eq:sumF} in $(a)$. It is readily verified that the choice of optimization and dual variables satisfy the KKT conditions and are hence optimal since the problem is convex \cite[Ch.~5.5.3]{Boyd2009}. Therefore, static caching maximizes \eqref{obj:single}.
\end{IEEEproof}
It remains to optimize over the $C_i$, but since, by Lemma~\ref{lem:kkt}, static caching is optimal for any assignment of $C_i$'s, it is optimal for \eqref{obj:sbs}--\eqref{cnstr:xi2sbs}.


\section{Proof of Theorem~\ref{th:fracknap}}
\label{prf:fracknap}
Letting $\gamma_1 = 1$, $\updatecost = 0$, and $\mbscost > \sbscost$, the STTL problem, i.e., maximizing \eqref{obj:minsingle} under constraints \eqref{cnstr:cache} and \eqref{cnstr:mu}, is equivalent to
\begin{align}
	\underset{\mu_{i,j} \in \R}{\text{maximize}}~ & \sum_{i=1}^N \omega_i s_i \sum_{j=0}^K \mu_{i,j} F_{i,j},\\
	\text{subject to}~ & \sum_{i=1}^N \omega_i s_i \sum_{j=0}^K \mu_{i,j} A_{i,j} \le C,\\
		& 1 \ge \mu_{i,0} \ge \mu_{i,1} \ge \cdots \ge \mu_{i,K} \ge 0.\label{cnstr:mufracknap}
\end{align}
Relaxing the constraint \eqref{cnstr:mufracknap}, replacing it with $0 \le \mu_{i,j} \le 1$, and letting $x_{i,j} = \omega_i s_i \mu_{i,j} A_{i,j}$, we obtain
\begin{align*}
	\underset{x_{i,j} \in \R}{\text{maximize}}~ & \sum_{i=1}^N \sum_{j=0}^K x_{i,j} \frac{F_{i,j}}{A_{i,j}},\\
	\text{subject to}~ & \sum_{i=1}^N \sum_{j=0}^K x_{i,j} \le C,\\
		& 0 \le x_{i,j} \le \omega_i s_i A_{i,j},
\end{align*}
which is recognized as the fractional knapsack problem \cite{Danzig1957}. The optimal solution to this problem is obtained by setting $x_{i,j} = \omega_i s_i A_{i,j}$, i.e., $\mu_{i,j} = 1$, greedily with respect to the fractions $F_{i,j}/A_{i,j}$ \cite{Danzig1957}. We observe that, since $F_{i,j}/A_{i,j}$ is decreasing in $j$ as explained in Sec.~\ref{sec:prel}, the constraint \eqref{cnstr:mufracknap} is met and we have a valid STTL caching policy. Apart from the possibility that one $\mu_{i,j}<1$ depending on the value of $C$, the constraints \eqref{cnstr:nu}--\eqref{cnstr:beta2} are also satisfied by this policy and, hence, the optimal STTL caching policy is equivalent to the optimal FTTL and TTL caching policies.
\end{appendices}


\bibliographystyle{IEEEtran}

\end{document}